\documentclass{llncs}
\usepackage{graphicx}
\usepackage{subfig}
\usepackage{multirow}
\usepackage{multicol}
\usepackage{booktabs}
\usepackage{algorithm}
\usepackage[noend]{algpseudocode}
\usepackage{amsmath}
\usepackage{arabtex}
\usepackage{utf8}
\usepackage[hyphens]{url}

\usepackage{color}
 \usepackage{xcolor}
 \usepackage{dingbat}
\usepackage{makecell}

\newcommand{\rowname}[1]
{\rotatebox{90}{\makebox[.3\linewidth][c]{\textbf{#1}}}}

\newcommand\ignore[1]{}

\usepackage{hyperref}

\begin{document}

\title{Malicious and Low Credibility URLs on Twitter during 
the AstraZeneca COVID-19 Vaccine Development}

\titlerunning{Malicious and Low Credibility URLs on Twitter during
COVID-19}
%
\author{Sameera Horawalavithana\inst{1}\orcidID{0000-0002-0327-3819} \and
Ravindu De Silva\inst{2} \and
Mohamed Nabeel\inst{3} \and
Charitha Elvitigala\inst{4} \and
Primal Wijesekara\inst{5} \and
Adriana Iamnitchi\inst{1}\orcidID{0000-0002-2397-8963}}
\authorrunning{S. Horawalavithana et al.}

\institute{University of South Florida
\\
\email{sameera1@usf.edu,anda@cse.usf.edu}
\and
SCoRe Lab
\\
\email{\{ravindud,charitha\}@scorelab.org}
\and
Qatar Computing Research Institute
\\
\email{mnabeel@hbku.edu.qa}
\and
University of California, Berkeley
\\
\email{primal@berkeley.edu}
}
\maketitle     

\begin{abstract}
We investigate the link sharing behavior of Twitter users  
following the temporary halt of AstraZeneca COVID-19 vaccine development in September 2020.
During this period, we show
the presence of malicious and low credibility information sources shared on Twitter messages in multiple languages.
The malicious URLs, often in shortened forms, are increasingly hosted in content delivery networks and shared cloud hosting infrastructures not only to improve reach but also to avoid being detected and blocked.
There are potential signs of coordination to promote both malicious and low credibility URLs on Twitter. 
Our findings suggest the need to develop a system that
monitors the low-quality URLs shared in times of crisis.
\end{abstract}


\section{Introduction}

During the COVID-19 pandemic, Twitter is being used to spread both high-quality and low-quality information~\cite{singh2020understanding}.
While information regarding the health, diseases, and vaccines are rapidly shared during this crisis period, there is a high prevalence of health misinformation that often contradicts with the opinions provided by health experts.
Many researchers called this an "infodemic" during COVID-19 pandemic~\cite{bagherpour_2020} that highlights the “safety, efficacy and necessity” concerns around vaccines~\cite{rory2020vaccinefirstdraft}.
Many factors contribute to the spread of health misinformation.
A recent report suggested that 20\% health misinformation claims are shared by politicians, celebrities and other prominent figures which received 69\% total social media engagement~\cite{bagherpour_2020}.
Both social media companies and public health officials are heavily criticized due to the delayed actions of controlling the circulation of these misleading messages~\cite{rory2020vaccinefirstdraft}.

There have also been active efforts to sow seeds of doubt on the efficacy of vaccines in the long term.
According to Reuters~\cite{reuters_2021}, AstraZeneza vaccine development faces many challenges from the date of its inception.
For example, there is a temporary halt in the development of AstraZeneca vaccine on September 2020 due to an unexplained illness that was reported in one of the trial participants~\cite{robbins_feuerstein_branswell_2020}.
AstraZeneca did not release enough details of this event which lead scientists to question its transparency on the vaccine development efforts~\cite{cyranoski2020scientists}.
This event also led The speaker of the House of Representatives, Nancy Pelosi to make a public statement regarding the approval of COVID-19 vaccines relying on UK safety tests~\cite{guardian_2020}.
Digital media shared various information about this event due to the popularity of AstraZeneca vaccine during the final stages of clinical testing~\cite{cyranoski2020scientists,guardian_2020}.

In this paper, we attempt to uncover the patterns of link sharing behavior on Twitter discussions following this event.
We discover a strong presence of malicious and low credibility information sources shared on Twitter messages in multiple languages.
We also show potential signs of coordination to promote these low-quality information.

\section{Data Collection and Processing}
We used a publicly available Twitter dataset around AstraZeneca COVID-19 vaccine development released as a part of Grand Challenge, North American Social Network Conference, 2021~\cite{nasn_2021}.
The keywords used to collect this dataset are~\emph{
        AstraZeneca,
Astra Zeneca,
AZD1222,
COVID,
vaccine,
immunity,
herd immunity,
Barrington,
and 
focused protection.
}

\begin{figure*}[htbp]
    \centering
    \includegraphics[scale=0.55]{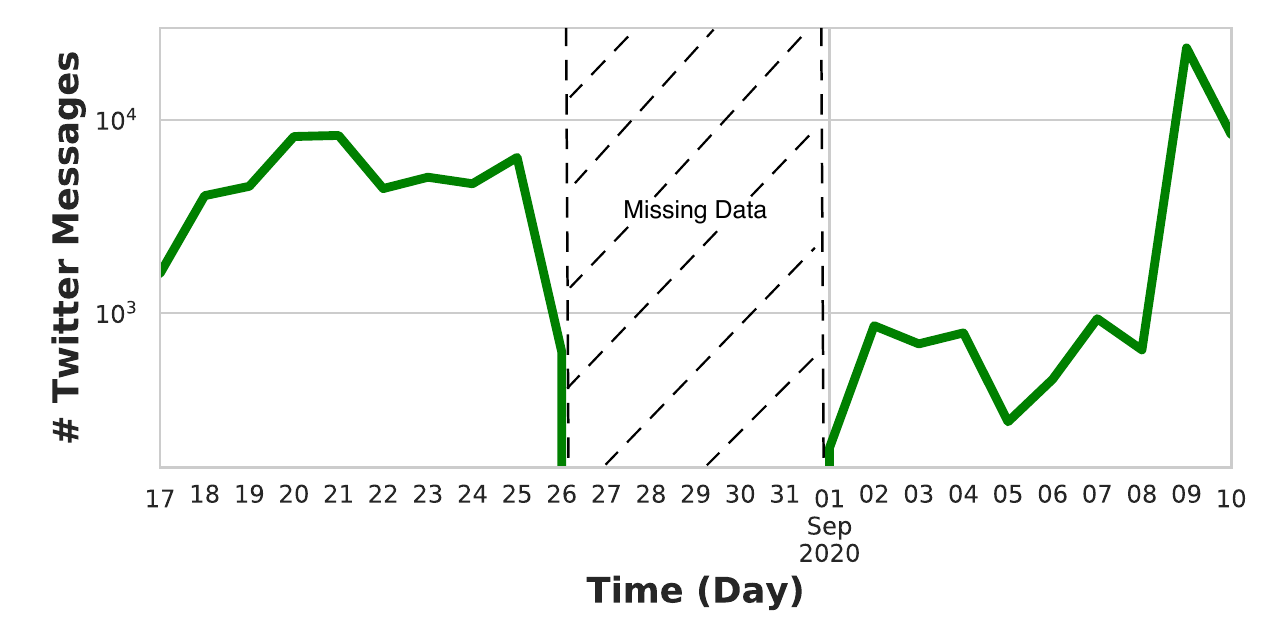}
    \caption{Number of Twitter URL mentions over time. There is a known gap in the original data collection between August 27-31~\cite{nasn_2021}}
    \label{fig:tweets_url_timeseries}
\end{figure*}

To understand the patterns of sharing \emph{Uniform Resource Locator} (URL) information in Twitter conversations, we only consider the Twitter messages that contain at least a URL, which consist of 25\% from the total messages.
As shown in Figure~\ref{fig:tweets_url_timeseries}, they are shared between August 17 and September 10, 2020.
Note that the original Twitter dataset covered a time period between August 17 and October 23, 2020, but include many gaps in the data after September 10, 2020.
We filtered out URLs to Twitter itself that typically refer to other tweets.
In addition, the external links (e.g., a tweet mentioning a YouTube video, or an external website domain) mentioned in the messages are pre-processed as following.
The shortened URLs are expanded, and HTML parameters are removed from the URLs.
The YouTube URLs are resolved to the base video URL if they include a parameter referencing a specific time in the video.
We represent the URL by the parent domain when multiple child domains exist (e.g., \texttt{fr.sputniknews.com}, \texttt{arabic.sputniknews.com} etc. are renamed as \texttt{sputniknews.com}).
This pre-processing code of resolving URLs is publicly available~\cite{PNNL}.
In this analysis, we do not differentiate among different message types (e.g., tweets, retweets, replies and quotes), but consider all messages as URL mentions.
The resulting dataset consists of 85,064 Twitter messages.
These messages are shared by 40,287 users citing 26,422 distinct URLs.
These URLs span over 6,990 distinct web domains.

Figure~\ref{fig:tweets_url_timeseries} shows a spike in URL mentions on September 9, 2020.
There are 23,620 (28\%) messages citing distinct 8,566 URLs on this particular day.
The most popular URLs point mainstream news articles, while the article published in \texttt{statnews.com} received the highest number (1,158) of shares~\cite{robbins_feuerstein_branswell_2020}.
This article reports the halt in AstraZeneca vaccine development following a suspected adverse reaction of a trial participant.

For additional analysis, we scrape the web page content pointed to an URL using a Python library, Newspaper.
We managed to scrape 22,856 articles. 
The tool failed to extract content from some web domains mainly due to inactive web pages and regulations enforced by the web domain.
According to the Langid~\cite{lui2012langid}, the majority of the articles are in English (6,453) and Spanish (5,197) languages.
Other articles are written in Turkish (1,839), French (1,539), Portuguese (1,277), Russian (784), Italian (772), Greek (708), Japanese (702), Croatian (447), etc.
There are 274 and 165 YouTube videos shared on Twitter have the title written in English and Spanish languages, respectively.

\section{Low Credibility Information Sources}
\label{sec:low_credible}

\begin{table}[htbp]
    \centering
        \caption{Twitter sharing characteristics for low credibility URLs as identified by MBFC and NewsGuard (NG). We use the~\checkmark~mark to reflect whether the domain is being listed as a low credibility information source by the respective fact checking organization.}
    \label{tab:basic_char_mbfc}
    \scalebox{0.8}{
    \begin{tabular}{|c|c|c|r|r|r|}
    \hline
         Domain &  MBFC & NG & \# Mentions & \# Users & \# URLs \\\hline \hline
 sputniknews.com & \checkmark & \checkmark  & 22,251 & 4,358 & 6,638 \\ \hline
rt.com & \checkmark & \checkmark  & 3,225 & 1,922 & 119 \\ \hline
zenith.news & \checkmark & - & 117 & 6 & 62 \\ \hline
zerohedge.com & \checkmark & \checkmark  & 278 & 249 & 15 \\ \hline
swarajyamag.com & \checkmark & - & 228 & 105 & 5 \\ \hline
oann.com & \checkmark & \checkmark  & 50 & 48 & 4 \\ \hline
childrenshealthdefense.org & \checkmark & \checkmark  & 3 & 3 & 3 \\ \hline
globalresearch.ca & \checkmark & \checkmark  & 16 & 7 & 3 \\ \hline
torontotoday.net & \checkmark & - & 9 & 5 & 2 \\ \hline
gnews.org & \checkmark & \checkmark  & 4 & 4 & 2 \\ \hline
truepundit.com & \checkmark & \checkmark  & 29 & 29 & 1 \\ \hline
needtoknow.news & \checkmark & - & 5 & 5 & 1 \\ \hline
oye.news & \checkmark & \checkmark  & 1 & 1 & 1 \\ \hline
gellerreport.com & \checkmark & \checkmark  & 1 & 1 & 1 \\ \hline
barenakedislam.com & \checkmark & - & 1 & 1 & 1 \\ \hline
wakingtimes.com & \checkmark & \checkmark  & 1 & 1 & 1 \\ \hline
thegatewaypundit.com & \checkmark & \checkmark  & 1 & 1 & 1 \\ \hline
    \end{tabular}
    }
\end{table}

\begin{figure*}[htbp]
\centering
\begin{tabular}{cc}
     \subfloat[Top-10 shared domains]{
    \includegraphics[width=0.42\linewidth]{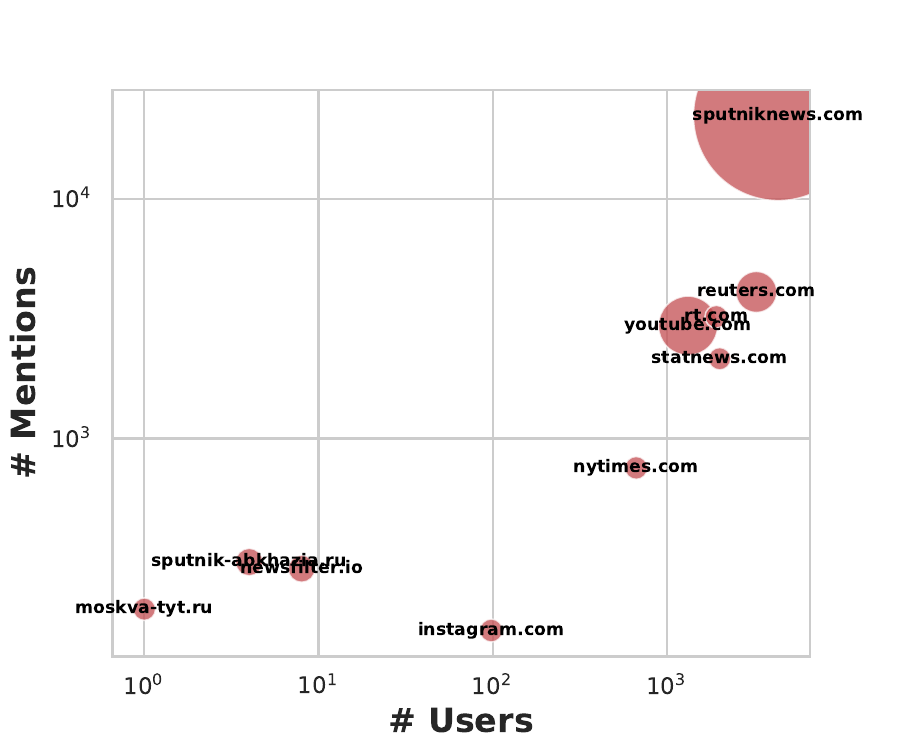}
    \label{fig:top10_domain_info}
    }
   &
     \subfloat[Twitter lifespan of URLs by domain]{
     \includegraphics[width=0.58\linewidth]{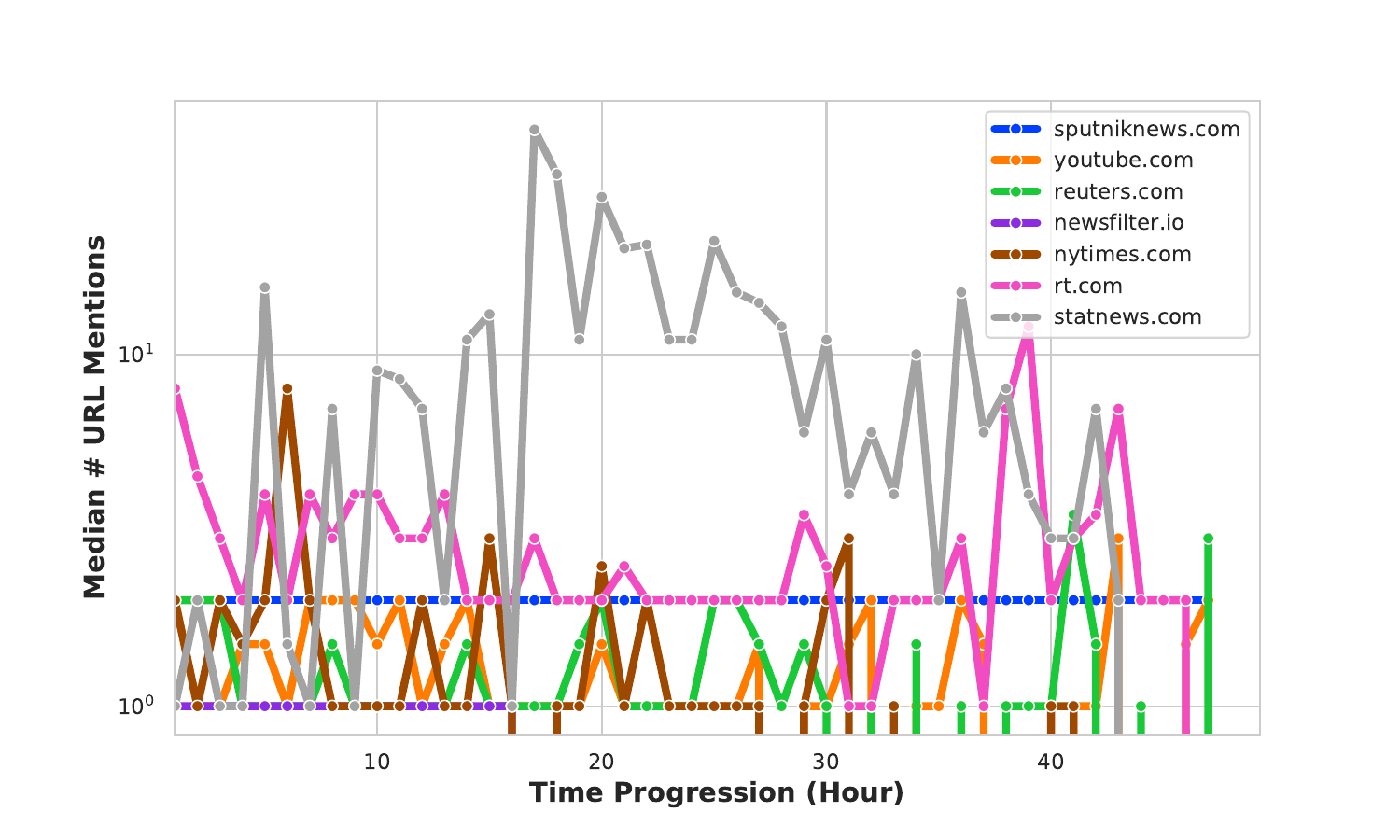}
     \label{fig:top10_domain_growth}
     }
     \\
\end{tabular}
	 \caption{Twitter sharing characteristics of most popular domains. Figure a) shows the Top-10 domains by the number of distinct URLs. The size of the markers in this plot are proportional to the number of URLs  associated  with  the  domain. Figure b) shows the number of URL mentions posted at each hour after the first appearance of URL in the respective domain. We count the number of mentions for each URL in the respective hour, and calculate the median number of mentions for the domain of the URL.}
	 \label{fig:top10_domain}
\end{figure*}

In this section, we report how Twitter users react to low credibility information sources.
We group the web domains according to the classification made by Media Bias/Fact Check (MBFC)~\cite{mbfc}.
We also cross-check these domains with the list of web sites that publish False COVID-19 information as identified by NewsGuard~\cite{newsguard_2021}.
We identify 6,860 (26\%) URLs from 17 low credibility information sources that are shared on 26,220 (31\%) messages (as shown in Table~\ref{tab:basic_char_mbfc}).
\texttt{sputniknews.com} is the most popular web domain by the number of mentions (22,251), the number of users (4,358) and the number of URLs (6,638).
This can be expected mainly due to a network of \texttt{sputniknews.com} media outlets that publish articles in many languages.
For example, 1,377 and 1,080 \texttt{sputniknews.com} articles are published in Turkish and French.
However, \texttt{sputniknews.com} URLs are not the most popular to be shared in the immediate hours after their first appearance (as shown in Figure~\ref{fig:top10_domain_growth}).
For example, the median number of \texttt{sputniknews.com} URL mentions is 2 in an hour.
\texttt{sputniknews.com} is frequently described as a Russian propaganda outlet that spreads master narratives in the Russia's disinformation campaign~\cite{mbfc}.

We noticed that \texttt{rt.com} acquires its many mentions from very few URLs: 119 URLs are cited in 3,225 messages during our observation period (Figure~\ref{fig:top10_domain_info}).
\texttt{rt.com} URLs have the highest number of mentions in the first hour after publication (as shown in Figure~\ref{fig:top10_domain_growth}) relative to the rest of popular web sites.

We also note that Twitter messages share the same article heading in the messages when they cite the same URL.
These users promoted certain topics through massive repetition of messages via injecting URLs.
For example, an article published in \texttt{zerohedge.com} was in the Top-10 most popular URLs on the day when AstraZeneza vaccine development halted~\footnote{https://www.zerohedge.com/markets/ft-confirms-astrazeneca-covid-19-vaccine-caused-serious-spinal-issues-test-patient}.
However, this article tried to build an alternative frame highlighting a statement by the US House Speaker Nancy Pelosi about the issue instead of reporting the details of the main event.
We also discovered URLs from two Russian news media domains, \texttt{sputnik-abkhazia.ru} and \texttt{moskva-tyt.ru} that are shared by very few users.
For example, \texttt{sputnik-abkhazia.ru} is the fourth most popular web domain by the number of URLs cited on Twitter messages.
These URLs are shared solely by the Twitter account controlled by the news media site.
However, they gain limited engagement with other users.

\begin{table*}[htbp]
	\centering
	\caption{The most popular domains hosting articles written in multiple languages. \texttt{sputniknews.com} and \texttt{youtube.com} are among the most popular domains.}
	\label{tab:lang_domain}
	\begin{tabular}{|c|}
		\hline
		English  \\ \hline \hline
		\textbf{sputniknews.com} \\ \hline
		\textbf{youtube.com} \\ \hline
		reuters.com \\ \hline
		newsfilter.io \\ \hline
		nytimes.com \\ \hline
	\end{tabular}
	\begin{tabular}{|c|}
		\hline
		Spanish  \\ \hline \hline
		\textbf{sputniknews.com} \\ \hline
		\textbf{youtube.com} \\ \hline
		rt.com \\ \hline
		reuters.com \\ \hline
		wp.me \\ \hline
	\end{tabular}
	\begin{tabular}{|c|}
		\hline
		Turkish  \\ \hline \hline
		\textbf{sputniknews.com} \\ \hline
		\textbf{youtube.com} \\ \hline
		ntv.com.tr \\ \hline
		is.gd \\ \hline
		sozcu.com.tr \\ \hline
	\end{tabular}
	\begin{tabular}{|c|}
		\hline
		French  \\ \hline \hline
		\textbf{sputniknews.com} \\ \hline
		cvitrolles.wordpress.com \\ \hline
		\textbf{youtube.com} \\ \hline
		francetvinfo.fr \\ \hline
		lalibre.be \\ \hline
	\end{tabular}
	\begin{tabular}{|c|}
		\hline
		Portugese  \\ \hline \hline
		\textbf{sputniknews.com} \\ \hline
		brasil247.com \\ \hline
		evsarteblog.wordpress.com \\ \hline
		\textbf{youtube.com} \\ \hline
		tvi24.iol.pt \\ \hline
	\end{tabular}
\end{table*}
\section{Malicious URLs}
\label{sec:malicious}
We use VirusTotal (VT) \cite{VirusTotal} to extract the \textit{maliciousness} of URLs.
VT provides the state-of-the-art aggregated intelligence for domains and URLs, and relies on more than 70 third-party updated antivirus (AV) engines.
 For all distinct URLs in our collection, we extract VT scan  reports via querying the publicly available API.
Each VT scan report contains of the verdict from every AV engine, information related to the URL such as first and last seen dates of the URL in the VT system, hosting IP address,  final redirected URL (if applicable), content length, etc.  
Each AV engine in a VT report detects if the URL is malicious or not. 
We use the number of engines that detect a URL as malicious as an indication of the maliciousness of the URL.

In this study, we label a URL as malicious if at least one AV engine, that is \#VT $\ge$ 1, detects it as malicious. Such malicious URLs, in general, are either phishing websites that steal user credentials and/or personally identifiable information from victims or malware hosting websites that attempt to install malware on victims' devices. We identify 441 
malicious URLs from the collected VT reports. We observe that 25.75\% of the malicious URLs utilize URL shortening services with top 4 services being \texttt{bit.ly}, \texttt{tinyurl.com}, \texttt{ow.ly} and \texttt{goo.su} whereas as only 0.97\% of benign URLs utilize such services.  This observation is consistent with the trend that malicious actors increasingly utilize shortening services to camouflage malicious URLs presenting non-suspecting URLs to users~\cite{fas2020malware}. 
We find that 30.80\% of the domains related to malicious URLs are ranked below 100K by Alexa~\cite{alexa} (The lower the rank value, the higher the popularity). This indicates the concerning fact that malicious actors are able to reach a large user base reaping a high return on investment for their attacks.

\begin{table}[ht]
\begin{minipage}[b]{0.55\linewidth}
\centering
\begin{tabular}{|l|l|}
\hline
\textbf{Feature}   & \textbf{Description}                                                                             \\ \hline \hline
VT\_Dur    & URL duration in VT                                    \\ \hline
PDNS\_Dur & Domain duration in PDNS \\ \hline 
\#IPs & \# hosting IPs \\ \hline 
\#Queries & \# times the domain is accessed \\ \hline
\#NSes & \# Name servers \\ \hline
Is\_NS & NS domain matches? \\ \hline
\#SOAs & \# administrative domains \\ \hline
Is\_SOA & Admin domain matches? \\ \hline
\#Domains & \# domains hosted on the IP \\ \hline
\#Queries\_IP & \# times the IP is accessed \\ \hline
ASN & Autonomous System Number \\ \hline
Org & Organization owning the ASN \\ \hline
\end{tabular}
    \caption{Details of the hosting features for malicious URL clustering}
    \label{tab:infra_features}
\end{minipage}\hspace{0.75mm}\hfill
\begin{minipage}[b]{0.45\linewidth}
\centering
\includegraphics[width=0.98\linewidth]{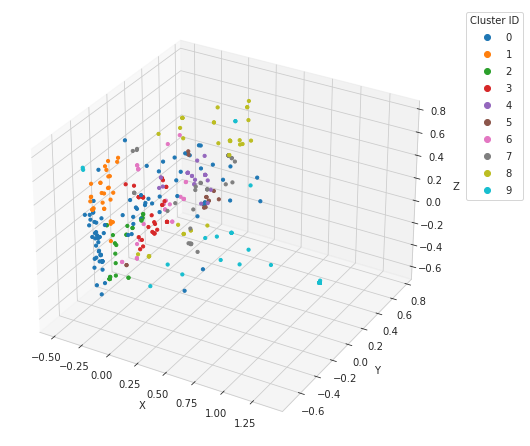}
\captionof{figure}{Malicious URL clusters based on the lexical and hosting features. Each point is a URL, and it is colored according to the cluster it belongs.}
\label{fig:t-sne-all}
\end{minipage}
\end{table}

We further analyze the malicious URLs to identify related malicious URLs. To this end, based on the lexical features in the literature~\cite{ravindu2021compromised} and the hosting features mentioned in Table~\ref{tab:infra_features}, we cluster the malicious URLs using PCA/OPTICS algorithm.

While lexical features identify characteristics related to URLs themselves, hosting features, extracted from Farsight Passive DNS (PDNS) data~\cite{dnsdb}, capture the characteristics of underlying hosting infrastructure. As shown in Figure~\ref{fig:t-sne-all}, these features collectively identify 10 distinct malicious URL clusters. We manually verified the accuracy of the top 2 clusters by checking the web
page content, registration information and domain certificate information. 
Our observations suggest that attackers launch multiple attacks at the same time.
We also analyze the clusters based on the maliciousness of URLs.
The maliciousness of a URL can loosely be measured by \#VT, the number of VT positives. An interesting observation is that URLs belonging to different maliciousness levels share similar lexical and hosting features.   
We further analyze these malicious URLs in terms of where they are hosted. To our surprise, we find that 80.04\% of these malicious URLs are hosted in CDNs such as Cloudflare and Akamai. While CDNs provide fast delivery of content across the globe through their distributed computing infrastructure, we believe a key reason why malicious actors utilize such services is to improve attack agility and stay below the radar of malicious domain detection mechanisms in place. This observation is further reinforced with the increased utilization of public cloud computing infrastructure (33.5\% of all malicious URLs) sharing hosting IPs with tens of thousands of unrelated domains, which are mostly benign. Such shared IPs are usually not blocked in practice due to the collateral damage.

\section{User Co-sharing Practices}
\label{sec:coordination}

As reported previously~\cite{pacheco2020uncovering}, there are signs of coordination to spread the disinformation content.
In this work, one of our assumptions is that the coordination is based on the content being shared (i.e., URLs mentioned in the tweets).
One of our objectives is to measure the extent of coordination in URL sharing activities.
We construct two bipartite networks to quantify the amount of such coordination effort.
The first network connects an author to a low credibility URL mentioned in a tweet, and the second network connects an author to a malicious URL mentioned in a tweet.
These networks would capture any suspicious behavior of promoting a particular URL.

We identify 6,860 low credibility information URLs that are shared by 6,600 users in 26,220 Twitter messages.
The five (0.08\%) most active users cite low credibility URLs in 6,404 (24\%) messages.
Their most preferred information source is 
the Turkish outlet of the sputnik media network.

We identify 441
malicious URLs that are shared by 357 users in 571 Twitter messages.
While a user might share a malicious URL unwittingly, it is suspicious to note users who share multiple malicious URLs.
Specifically, we identify 51 users who share more than 1 malicious URL, and 21 users who share more than 2 malicious URLs.

\begin{figure*}[htbp]
\centering
\begin{tabular}{cc}
     \subfloat[Coordination to promote low credibility URLs (KS=0.6, p-value=0.001)]{
    \includegraphics[width=0.48\linewidth]{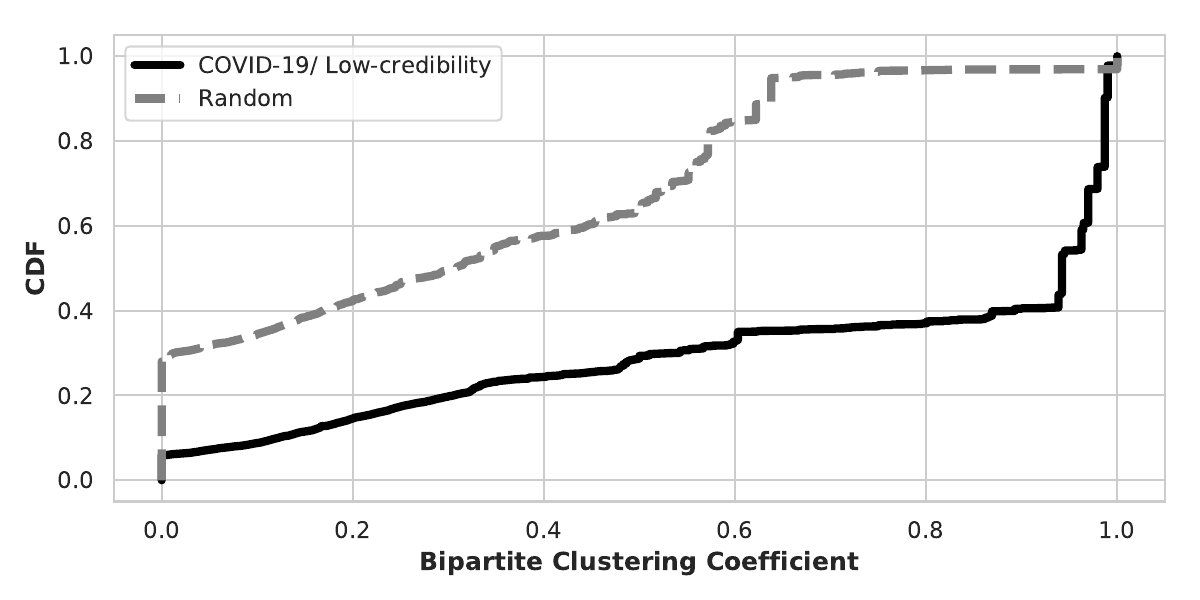}
    \label{fig:bipartite_clustering_user_mbfc_url}
    }
   &
     \subfloat[Coordination to promote malicious URLs (KS=0.27, p-value=4.45$e^{-12}$)]{
     \includegraphics[width=0.48\linewidth]{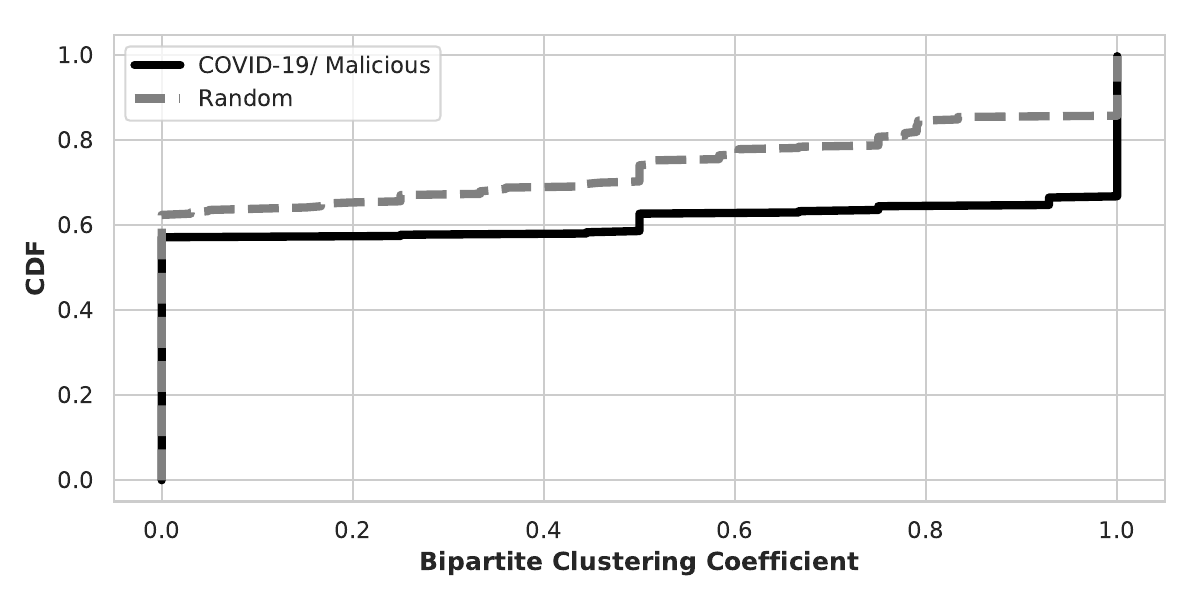}
     \label{fig:bipartite_clustering_user_mal_url}
     }
     \\
\end{tabular}
	 \caption{Bipartite clustering coefficient of a) the user-(low credibility) URL network, b) the user-(malicious) URL network. We also compare the network clustering values with a random bipartite network using the Newman's configuration model~\cite{newman2003structure}. The deviation of the clustering values from the random bipartite network shows potential coordination effort to promote these URLs on Twitter.}
	 \label{fig:bipartite_clustering}
\end{figure*}

Figures~\ref{fig:bipartite_clustering_user_mbfc_url} and~\ref{fig:bipartite_clustering_user_mal_url} show the distributions of bipartite clustering coefficients~\cite{latapy2008basic} for URLs in the two bipartite networks respectively.
Clustering coefficient values are higher for URLs when they are shared by a group of users who engage with other URLs together.
We compare the similar clustering values which are calculated from the identical random bipartite networks.
We construct two random bipartite networks using the Newman's configuration model~\cite{newman2003structure} for the comparison.
Given the original user-URL bipartite network, we match the two degree sequences in the users and URLs in the random bipartite network.
We noticed a significant deviation of clustering coefficient values for URLs in both bipartite networks compared to URLs in the random bipartite networks (as shown in Figures~\ref{fig:bipartite_clustering_user_mbfc_url} and~\ref{fig:bipartite_clustering_user_mal_url}).
For example, there are 4,320 (63\%) low credibility URLs with a clustering coefficient value greater than 0.8 than those number of URLs (217) in the random bipartite network.
In the malicious URL network, there are 121 URLs with a clustering coefficient value greater than 0.8 that are promoted by the same set of users (the expected number of URLs is 36 in the random network).
To confirm our observation, we also perform the Kolmogorov-Smirnov (KS) test between the clustering coefficient values from the original and random bipartite networks.
KS-statistic values (statistically significant) are 0.6 and 0.27 for the low credibility URL network and malicious URL network, respectively.
This suggests the sustained effort of users to amplify both misleading and malicious content.
There may be different types of users (e.g., bots, cyborgs, paid activists) who amplify these URLs.
That would remain as a future work to identify these types of users.

\section{Conclusions}

In times of crisis, whether political or health-related, online disinformation is amplified by social media promotion of alternative media outlets~\cite{horawalavithana2020twitter}. 
This study adds to the growing body of work~\cite{ferrara2020misinformation} that analyzes the misinformation activity during the COVID-19 crisis by studying the sharing of URLs on Twitter between August 17 and September 10, 2020. 
Our contributions complement previous observations~\cite{singh2020understanding,fas2020malware} in multiple ways. 
We discover a strong presence of malicious and low credibility information sources shared on Twitter messages in multiple languages.
Not only that URLs from low credibility sources, as classified by independent bodies such as NewsGuard and Media Bias/Fact Check, were present, but many were shown to point to pages with malicious code. 
A significant portion of these URLs (25\%) were in shortened form (compared to under 1\% of the non-malicious URLs) and hosted on well-established, reputable content delivery networks 
in an attempt, we believe, to avoid detection.

We also discovered potential signs of coordination to promote malicious and low credibility URLs on Twitter. 
Specifically, we discovered unusual clustering of user activity related to the sharing of such URLs. 
We use a null model and several statistical tests to compare expected behavior with what we suspect to be coordinated behavior as seen in this dataset.

In general, we unmask malicious strategies that exploit Twitter to promote shady objectives in times of crisis.
Further work is needed to understand these objectives.
For example, bad actors might have chosen this event strategically to maximize the spread of malicious URLs.
These actors can deploy the same strategy in the future conversations, thus having content moderation techniques to limit what they can share is important.
On the other hand, the low credibility news sources might have reported this event opportunistically in an attempt to promote vaccine hesitancy.
People might have engaged with these low quality sources to watch out the information space around COVID-19 vaccines.
According to Smith et al.~\cite{rory2020vaccinefirstdraft}, there is a deficit of high quality information sources to seek vaccine information.
Bad actors use this information deficit as an advantage to push low quality information.
We believe this analysis can be extended in understanding the role of bad actors during similar emotionally charged conversations in the future.

\subsubsection*{Acknowlegements}
This work is partially supported by the DARPA SocialSim Program and the Air Force Research Laboratory under contract FA8650-18-C-7825. 
The authors would like to thank Grand Challenge, North American Social Network Conference for providing data.

\small
\bibliographystyle{splncs04}
\bibliography{main}

\end{document}